\documentclass[preprint,12pt]{elsarticle}

\usepackage{amssymb}
\usepackage{graphicx}

\usepackage{scalerel}
\newcommand{\comment}[1]{}
\usepackage{physics}
\usepackage[lofdepth,lotdepth]{subfig}
\usepackage{booktabs}
\usepackage{color}

\journal{Physics Letters B}

\begin{document}

\begin{frontmatter}


\title{Coherence dynamics in low-energy nuclear fusion}
 \author{Iain Lee}
 \author{Alexis Diaz-Torres\corref{cor1}}
 \ead{a.diaztorres@surrey.ac.uk}
 \cortext[cor1]{corresponding author}


\affiliation{organization={Department of Physics, University of Surrey},
            addressline={Faculty of Engineering and Physical Sciences}, 
            city={Guildford},
            postcode={GU2 7XH}, 
            country={United Kingdom}}

\begin{abstract}

Low-energy nuclear fusion reactions have been described using a dynamical coupled-channels density matrix method, based on the theory of open quantum systems. For the first time, this has been combined with an energy projection method, permitting the calculation of energy resolved fusion probabilities. The results are benchmarked against calculations using stationary Schr\"{o}dinger dynamics and show excellent agreement. Calculations of entropy, energy dissipation and coherence were conducted, demonstrating the capability of this method. It is evident that the presence of quantum decoherence does not affect fusion probability. This framework provides a basis for quantum thermodynamic studies using thermal environments. 

\end{abstract}





\end{frontmatter}


\section{Introduction}
\label{introd}

In stars, nuclear fusion takes place within a plasma --- a hot ionised gas where atoms are stripped apart into electrons and positively charged nuclei. For fusion to occur, the Coulomb barrier between a projectile and target must be overcome or tunnelled through. The Coulomb barrier is formed from the action of two opposite nucleus-nucleus interactions: the short-range, attractive nuclear interaction and the long-range, repulsive Coulomb interaction. There is scarse literature detailing how stellar plasmas affect nuclear fusion reactions, apart from studies on plasma screening \cite{Yakovlev1989}. Interactions on the boundary of nuclear and atomic physics are theorised to excite the atomic nucleus, and in a dense plasma environment, these excitations may impact fusion reaction rates \cite{Harston1999,adriana2006,Gosselin2009}. We present a method for describing low-energy nuclear fusion of heavy ions, with the capability to include effects from environments that influence these fusion reactions. 



The coupled-channels model is combined with an open quantum systems approach to model nuclear fusion. An open quantum system involves an environment that interacts with the quantum system of interest (or \textit{reduced system}) \cite{breuer2002}. This is necessary to describe physical systems, and it is of great interest across disciplines \cite{rotter2015} since the medium in which these systems evolve is often neglected in calculations. In low-energy nuclear physics, the theory of open quantum systems has been applied to understanding various dissipative quantum phenomena, such as deep-inelastic heavy-ion collisions \cite{sandulescua1987} and heavy-ion capture reactions \cite{vazgen2016}. For nuclear fusion reactions, the inclusion of environments may lead to increased population of the intrinsic nuclear excited states, loss of quantum coherence (decoherence) and energy dissipation, therefore affecting the fusion probability \cite{Dasgupta2007}. As a result of decoherence and dissipation to an environment, we cannot work with pure states and instead we use mixed states. These require the use of a density matrix, instead of a wavefunction, to propagate the dynamics in time. To do this, we use a coupled-channels density matrix (CCDM) approach based on the Lindblad master equation \cite{Lindblad1976}. The CCDM approach was introduced in chemical physics \cite{pesce1997} and was later adapted to the field of nuclear physics, where it was used to investigate dissipative quantum dynamics in nuclear collisions \cite{alexis2008,alexis2010}. 
 
 
A key feature of this work is the ability to compare our dynamical (time-dependent) CCDM technique with stationary (time-independent) methods, such as the model implemented in the CCFULL code \cite{ccfull}, which uses the time-independent Schr\"{o}dinger equation (TISE). We do this by using the window operator \cite{schafer1991} as an energy spectral decomposition method to calculate energy-resolved fusion probabilities. This has been used in literature on wavefunctions \cite{boselli2015,alexis2018,terry2019}, but for the first time, we have applied the window operator to a final (asymptotic) density matrix. We also present pioneering work on the dynamics of quantum coherence and entropy production in nuclear fusion, as the CCDM approach allows us to study these important concepts of the emerging field of quantum thermodynamics \cite{kosloff2019}.

For the sake of simplicity and as a test case, we currently consider only a single internal vibrational excitation mode of the $^{144}$Sm target that interacts with an inert $^{16}$O projectile. This is the test reaction in the CCFULL code \cite{ccfull}. This paper serves as a proof of concept for the novel CCDM technique and the preliminary results provide a benchmark for this approach compared to the standard, time-independent coupled-channels method \cite{Balantekin1998,Hagino2012}. 



\section{Theoretical framework}
\label{section:formalism}

In the present model calculations, there is a projectile nucleus ($^{16}$O) and a target nucleus ($^{144}$Sm) that are initially separated at a large distance so that virtually no interactions between them are present (due to nuclear and Coulombic effects) and they reside in their ground states. The collision described is central and it is assumed that there is no change in orbital angular momentum $(L = 0)$. A ``fusion'' environment is used for describing dissipation out of the two-body direct nuclear reaction (reduced system).

\subsection{Reduced-system Hamiltonian, fusion environment and initial density matrix} 

\begin{figure}[htb!]
 \centering
 \includegraphics[scale=0.85]{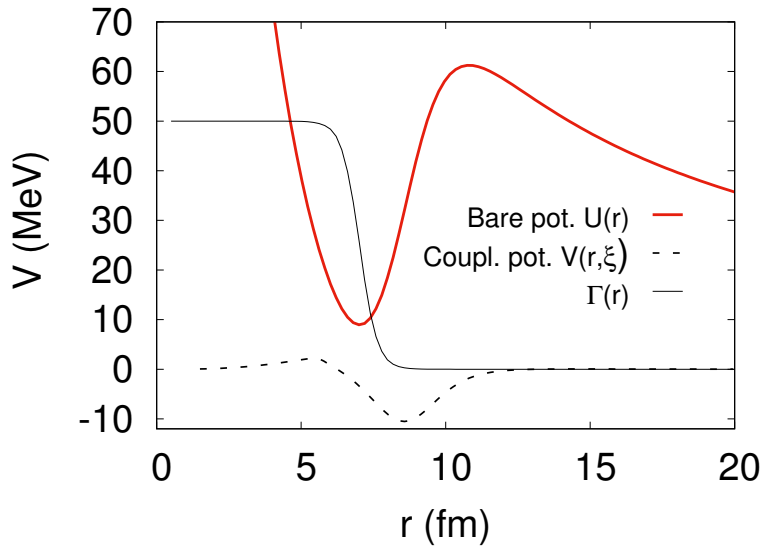}
 \caption{\label{fig:potential} The potentials for the test collision $^{16}$O $+$ $^{144}$Sm as a function of the internuclear radius. The thick red solid line is the total, bare nucleus-nucleus potential $U(r)$ and the dashed line is the total coupling potential, $V(r,\xi)$, between the ground-state and the $3^{-}$ vibrational excited state ($1.81$ MeV) of $^{144}$Sm. The thin black solid line is the decay function $\Gamma(r)$, which is a Fermi function that removes positional probability from the reduced-system density matrix due to compound nucleus formation.}
\end{figure}

The reduced-system Hamiltonian in the overall centre-of-mass reference frame, $\hat{H}_S$, is the same as the Hamiltonian implemented in the CCFULL code, and realistic parameters of the model can be found in Ref. \cite{ccfull}: 

\begin{equation}
    \hat{H}_{S} = \hat{T}(r) + \hat{U}(r) + \hat{h}(\xi) + \hat{V}(r,\xi),
\label{eq1}    
\end{equation}

where $\hat{T}(r)$ and $\hat{U}(r)$ are the kinetic energy operator and the monopole, total real interaction potential (nuclear + Coulomb parts) between the two nuclei in their ground states, respectively. $\hat{h}(\xi)$ is the intrinsic Hamiltonian of the nuclei, whose eigenstates and eigenenergies are $\ket{\alpha}$ and $e_{\alpha}$, respectively. The total real coupling potential, $\hat{V}(r,\xi)$, determines how the radial motion affects the population of the internal energy spectrum of the interacting nuclei. The model values of $U(r)$ and $V(r,\xi)$ for the $^{16}$O $+$ $^{144}$Sm collision are presented in Fig. \ref{fig:potential} to show the position and strength of these potentials. Plotting $V(r,\xi)$ shows the localisation of the coupling potential between the ground-state and the $3^{-}$ vibrational excited state ($1.81$ MeV) of $^{144}$Sm. This coupling is negligible until a position near the Coulomb barrier is reached.

The fusion channel has often been treated using a short-range imaginary potential in other coupled-channels fusion models, since it is not explicitly included in those model calculations \cite{thompson2001}. We include the fusion channel in our calculations,  and call it the \textit{fusion environment} which represents the high energy-density compound-nucleus states. One can imagine that the fusion environment acts like a detector, performing a continuous position measurement, effectively measuring the distance between two nuclei within a certain region of space where fusion takes place, causing loss of information and increasing entropy production. In this dynamical theory, we extend the Hilbert space of the system to include the fusion environment, which is modelled by an auxiliary state. This idea was first introduced in Ref. \cite{bertlmann2006} to describe particle decay. The coupling of the reduced system to the fusion environment is irreversible, and can be described by specific Lindblad operators related to an imaginary potential, $-i\Gamma(r)$ \cite{thompson2004}. The $\Gamma(r)$ function in Fig. \ref{fig:potential} determines the radial strength of the absorption to the fusion environment. By recording the information that leaves the reduced system with a fusion environment, we are able to quantify the irreversible effects on the collision dynamics due to energy dissipation, entropy production and loss of coherence.

To describe the initial radial motion of the nuclei, we have tested two different initial wave packets with the same Gaussian envelope but two different boosts, $\mathcal{B}(k_{0}r)$, namely a plane wave ($e^{-ik_{0}r}$) and an incoming Coulomb wave, $H^{-}_{L=0} (k_{0}r)$ \cite{thompson1986}: 

\begin{equation}
    \psi (r,r_{0},\sigma _{0},k_{0}) = \mathcal{N}^{-1} \, \exp [- \frac{(r-r_{0})^2}{2 \sigma _{0}^{2}}] \mathcal{B}(k_{0}r),
\label{eq2}    
\end{equation}

where $\mathcal{N}$ is a normalisation constant, $r_0$ is the initial, central position of the wave packet, $r$ is a radial grid position, $\sigma _0$ is the spatial dispersion, and $k_{0}$ is the average wave number, which depends on the average incident energy $E_0$, $r_0$ and $\sigma_0$ and is found by solving $E_0 = \bra{\psi}\hat{H}_S\ket{\psi}$. Eq. (\ref{eq2}) with a plane-wave boost is a Gaussian wave packet, while this is a Coulomb wave packet \cite{terry2021} when a Coulomb-wave boost is used. These wave packets were used to construct the radial part of the initial density matrix. 
  



 
The density matrix created from either of the wavefunctions above would be correct if the target and projectile nuclei were to remain only in their ground states. However, for every position on the radial grid, the target and/or projectile could be in an excited state. Therefore the density operator is a tensor and is formed from a mixed basis of states: the radial position and the intrinsic energy states, $\ket{r}$ and $\ket{\alpha}$, respectively:

\begin{equation}
    \hat{\rho} = \sum _{\alpha,\beta,r,s} \ket{r} \ket{\alpha} \rho ^{rs}_{\alpha \beta}(t) \bra{\beta} \bra{s}.
\label{eq3}    
\end{equation}

In Eq. (\ref{eq3}), the auxiliary state associated with the fusion environment is also included. The density operator must be Hermitian and positive semi-definite because the diagonal elements of the density matrix are the coefficients $\rho_{\alpha \alpha}^{rr}(t)$, which have the physical meaning of probabilities. Initially, when the nuclei are far apart, only the matrix elements of the density matrix related to the ground state of the nuclei are nonzero:

\begin{equation}
    \rho ^{rs}_{1 1}(t=0) = \ket{r} \, \bra{s}.
\label{eq4}    
\end{equation}

\subsection{Time propagation of the coupled-channels density matrix and the dynamics of quantum coherence}

Inserting Eq. (\ref{eq3}) into the Lindblad master equation for the density operator and projecting this equation onto both the radial grid position and intrinsic energy states, the equations of motion describing the time evolution of the coupled-channels density matrix, $\rho ^{rs}_{\alpha \beta}(t)$, are obtained \cite{alexis2008,alexis2010}. These are separated into matrix elements for the reduced system:

\begin{multline}
    \dot{\rho}^{rs}_{\alpha \beta} = - \frac{i}{\hslash} \left[ \rho ^{rs}_{\alpha \beta} \: (e_{\alpha} - e_{\beta}) + \sum ^{M}_{t=1} (T^{rt} \: \rho ^{ts}_{\alpha \beta} -  \rho ^{rt}_{\alpha \beta} \: T^{ts}) \right. \\
    +  \left. \rho ^{rs}_{\alpha \beta} \: (U^{rr} - U^{ss}) + \sum ^{N}_{\mu =1} (V^{rr}_{\alpha \mu} \: \rho ^{rs}_{\mu \beta} -  \rho ^{rs}_{\alpha \mu} \: V^{ss}_{\mu \beta})  \right]  \\
    + \delta _{\alpha \beta} \sum _{\nu =1}^{B} \sqrt{\Gamma ^{rr}_{\alpha \nu}} \, \rho ^{rs}_{\nu \nu} \sqrt{\Gamma ^{ss}_{\alpha \nu}} -\: \frac{1}{2} \sum _{\nu =1}^{B} (\Gamma ^{rr}_{\nu \alpha} + \Gamma ^{ss}_{\nu \beta}) \rho ^{rs}_{\alpha \beta},
\label{eq5}    
\end{multline}

and matrix elements involving the auxiliary state(s):

\begin{equation}
    \dot{\rho}^{rs}_{kl} = \delta _{kl} \sum _{\nu =1}^{B} \sqrt{\Gamma ^{rr}_{k\nu}} \, \rho ^{rs}_{\nu \nu} \sqrt{\Gamma ^{ss}_{k \nu}} - \frac{1}{2} \sum _{\nu =1}^{B} (\Gamma ^{rr}_{\nu k} + \Gamma ^{ss}_{\nu l}) \rho ^{rs}_{kl},
\label{eq6}
\end{equation}

where either $k$ or $l$ is equal to the index of any auxiliary state and $\Gamma^{rr}_{\nu \nu} = \sum_{\mu \ne \nu} \Gamma^{rr}_{\mu \nu}$ \cite{alexis2010}. The auxiliary state in Eq. (\ref{eq6}) allows one to store the information transferred out of the reduced system. In the present model calculations, $\Gamma^{rr}_{31}$ and $\Gamma^{rr}_{32}$ describe the irreversible coupling between the reduced system's states (denoted by 1 and 2) and the auxiliary state (denoted by 3). These matrix elements are equal to the function $\Gamma(r)$ in Fig. \ref{fig:potential}, whilst $\Gamma^{rr}_{21} = \Gamma^{rr}_{12} = 0$. In Eq. (\ref{eq5}), the coupling matrix $V^{rr}_{\alpha \beta} \equiv \bra{\alpha} V(r,\xi) \ket{\beta}$. Eqs. (\ref{eq5}) and (\ref{eq6}) are directly integrated using the Faber polynomial integrator \cite{huisinga1999} with the initial condition given by Eq. (\ref{eq4}) and a time step of $1 \times 10^{-22}$ s. The initial radial centroid of the wave packet was $r_0 = 70$ fm, and its typical spatial dispersion was $\sigma _0 = 10$ fm. The radial grid ($r = 1.5–150$ fm) was evenly spaced with $M = 1024$ points. The number of energy states of the reduced system was $N=2$, and the total number of intrinsic energy states was $B=3$.

The dynamics of quantum coherence in the reduced system can be described by the time-dependent purity, $\bar{\mathcal{P}}(t)$, and the von Neumann entropy, $\mathcal{S}(t)$, of the reduced-system density matrix, $\rho_S(t)\equiv \{\rho ^{rs}_{\alpha \beta}(t)\}$ from Eq. (\ref{eq5}):

\begin{equation}
\bar{\mathcal{P}}(t) = \textnormal{Tr} [\bar{\rho}_S^2(t)],
\label{eq_purity}
\end{equation}

\begin{equation}
\mathcal{S}(t) = - \textnormal{Tr} \left \{ \rho_S(t) \, \textnormal{ln} \,[\rho_S(t)] \right \} = - \sum_j \eta_j(t) \, \textnormal{ln} \, [\eta_j(t)],
\label{eq_entropy}
\end{equation}

where $\bar{\rho}_S(t)$ is the diagonal-removed reduced system density matrix \cite{Saalfrank1995,lenton2021} (i.e., $\bar{\rho}_S(t) \equiv \{\rho ^{rs}_{\alpha \beta}(t)\}$ with $\rho ^{rr}_{\alpha \alpha}(t)=0$), and $\eta_j(t)$ are the eigenvalues of $\rho_S(t)$ \cite{kosloff2019}. Before using these equations, the trace of $\rho_S(t)$ should be normalised to unity. A constant value of purity indicates a completely pure (coherent) state, showing that the interaction with the environment has not caused a mixed state. Conversely, a decrease of purity over time indicates decoherence caused by the effect of the environment. 

\subsection{Energy projection of the time-propagated density matrix for fusion probability calculations}

We model the initial projectile-target radial motion as a wave packet containing a distribution of energies, with average energy $E_{0}$. To calculate the fusion probability of the wave packet, the standard approach is to sum the fusion probability contributions from all energies within the wave packet, obtaining the total fusion probability for a fusion reaction. However, this causes the loss of small probabilities from energies well-below the Coulomb barrier, since the higher energies dominate the fusion probability. Therefore, a method is needed to resolve the energies of the wave packet and calculate energy-resolved fusion probabilities to fully understand these low-energy contributions. To this end, we employ the window operator, $\hat{\Delta}(E_{k}, n, \epsilon)$ \cite{schafer1991}:

\begin{equation}
    \hat{\Delta}(E_{k}, n, \epsilon) = \frac{\epsilon^{2^{n}}}{(\hat{H}_{S}-E_{k})^{2^{n}} + \epsilon ^{2^{n}}},
\label{eq7}    
\end{equation}

where $E_{k}$ is the incident energy of interest, $\epsilon$ is an energy resolution parameter that determines the width of the energy window, and $n$ is a parameter that determines the shape of the energy window. For a large $n$, the window function is increasingly rectangular.

The amount of probability of the reduced-system density matrix, $\rho_{S}$, in an energy window of width $2\epsilon$ around $E_{k}$ is:

\begin{equation}
    \mathcal{P}(E_{k}, n, \epsilon) = \textnormal{Tr} \big [ \hat{\Delta}(E_{k}, n, \epsilon) \, \rho_{S} \big ].
\label{eq8}    
\end{equation}

By expanding Eq.(\ref{eq8}) with $n=2$, the successive linear equations to obtain an energy-resolved density matrix, $\rho '_{k}$, can be written as: 

\begin{equation}
   (\hat{H}_{S} - E_k + i^{\frac{3}{2}} \epsilon) (\hat{H}_{S} - E_k - i^{\frac{3}{2}} \epsilon)  (\hat{H}_{S} - E_k + \sqrt{i} \epsilon) (\hat{H}_{S} - E_k - \sqrt{i} \epsilon) \rho '_{k} = \epsilon^4 \, \rho_{S},
\label{eq9}   
\end{equation}

and Eq. (\ref{eq8}) can be rewritten as:

\begin{equation}
    \mathcal{P}(E_{k}, n=2, \epsilon) = \text{Tr} ( \rho '_{k} ).
\label{eq10}    
\end{equation} 

Eqs. (\ref{eq9}) and (\ref{eq10}) have been used to calculate the fusion probability, $P_{fus}$, for a range of below-barrier to above-barrier incident energies, $E_k$:

\begin{equation}
    P_{fus}(E_{k}) = 1 - \frac{ \mathcal{P}_{final}(E_k)}{\mathcal{P}_{initial}(E_k)},
\label{eq11}    
\end{equation}

where $\epsilon$ was equal to $0.35$ MeV. In Eq. (\ref{eq11}), $\mathcal{P}_{initial}$ and  $\mathcal{P}_{final}$ refer to the energy spectrum associated with the initial and final reduced-system density matrices, respectively. Having applied the window operator, we can now compare our results of $P_{fus}$ from density matrix calculations with those from TISE calculations.

\section{Results}
\label{section:results}

\begin{figure}[ht!]
\centering
\subfloat[Subfigure 1 list of figures text][Elastic channel]{
\includegraphics[width=0.48\textwidth]{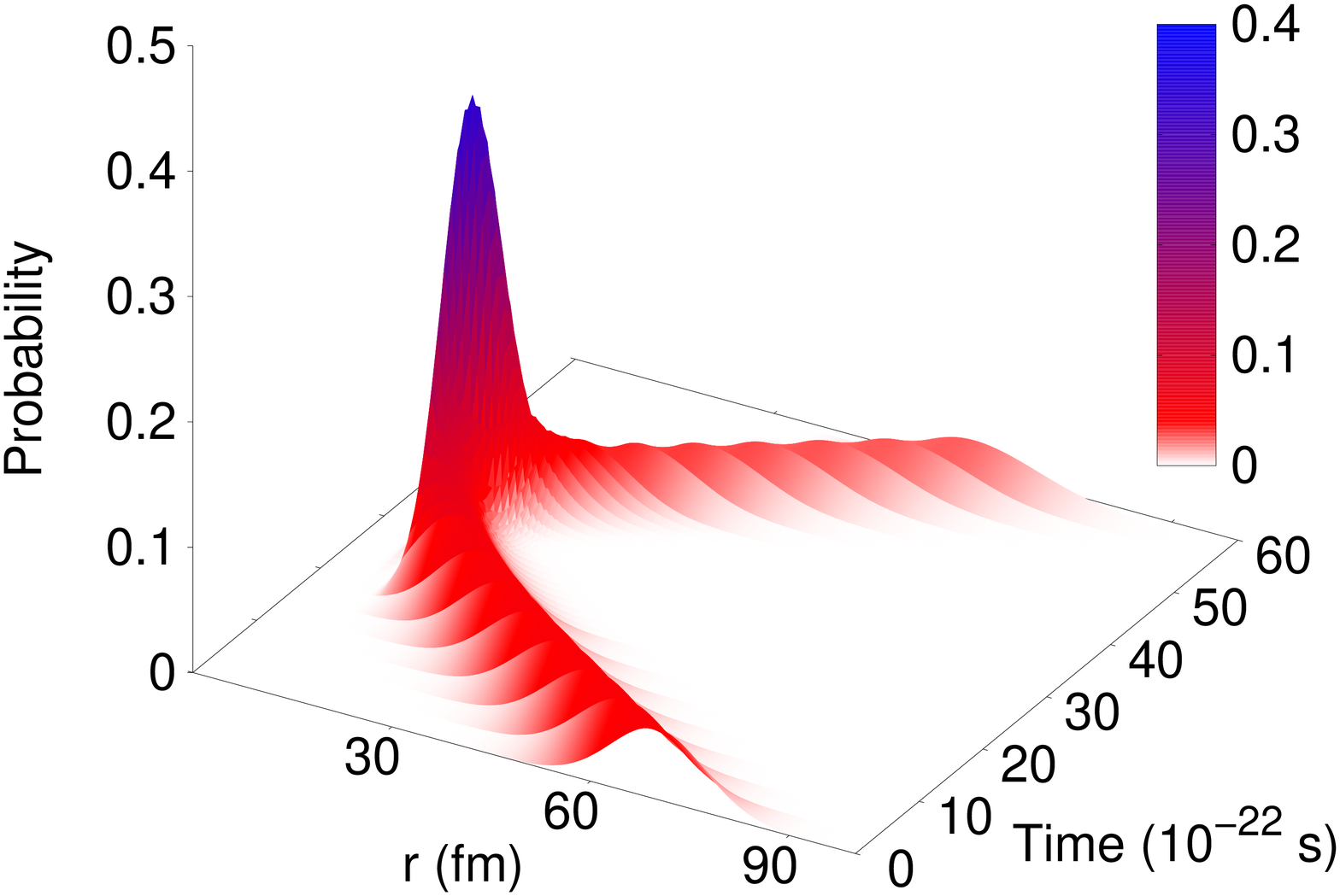}

\label{fig:subfig1}}
\subfloat[Subfigure 2 list of figures text][Inelastic channel]{
\includegraphics[width=0.48\textwidth]{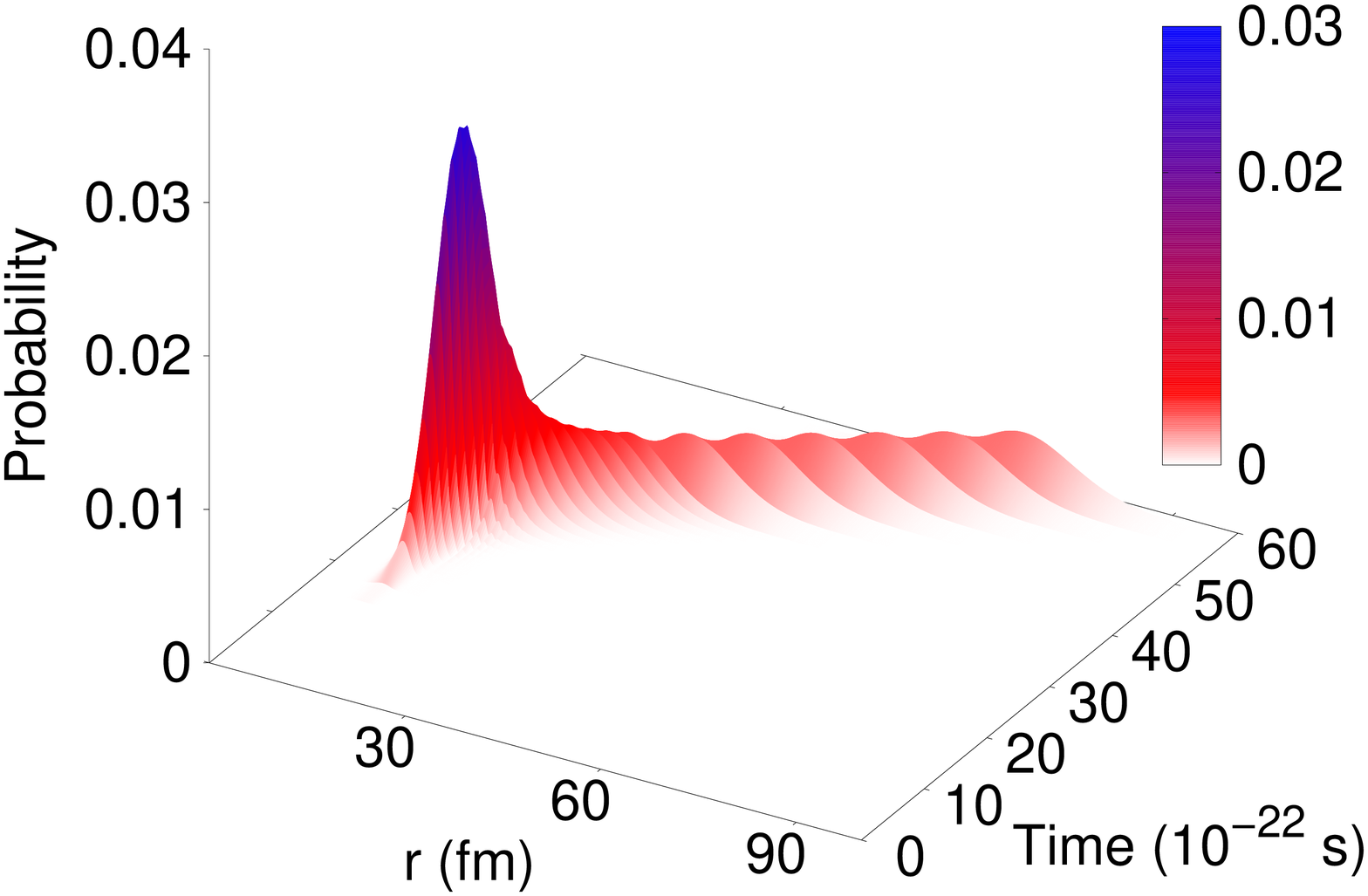}
\label{fig:subfig2}}
\caption{Radial position probability as a function of internuclear radius and time for a head-on collision of $^{16}$O + $^{144}$Sm with a mean energy of $60$ MeV. The radial probability decreases and increases for (a) the elastic and (b) inelastic channels respectively, as the nuclei approach their Coulomb barrier ($r \approx 10$ fm). 
For visualisation, when the mean radius is larger than $20$ fm, the time step is $3 \times 10^{-22}$ s.}
\label{fig:globfig}
\end{figure}

\begin{figure}[htb!]
 \centering
\subfloat{
\includegraphics[width=0.5\textwidth]{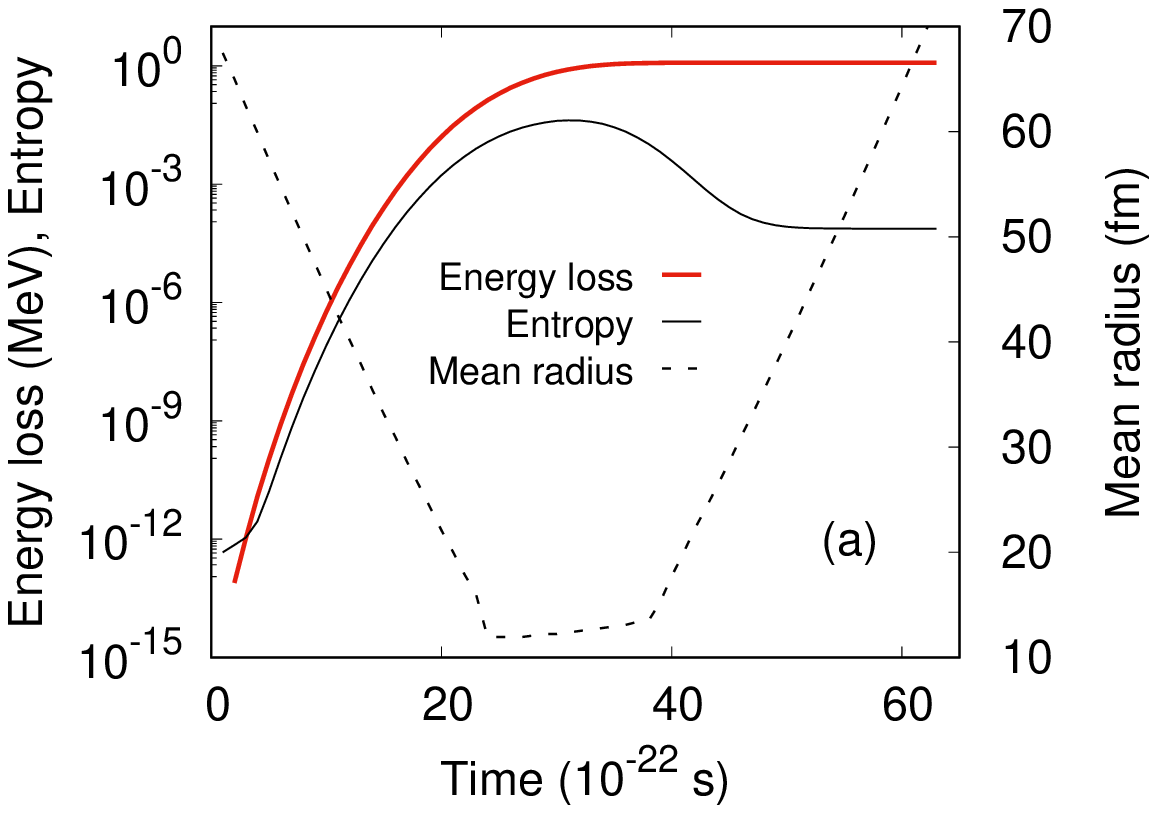}
\label{fig:coherence}}
\subfloat{
\includegraphics[width=0.5\textwidth]{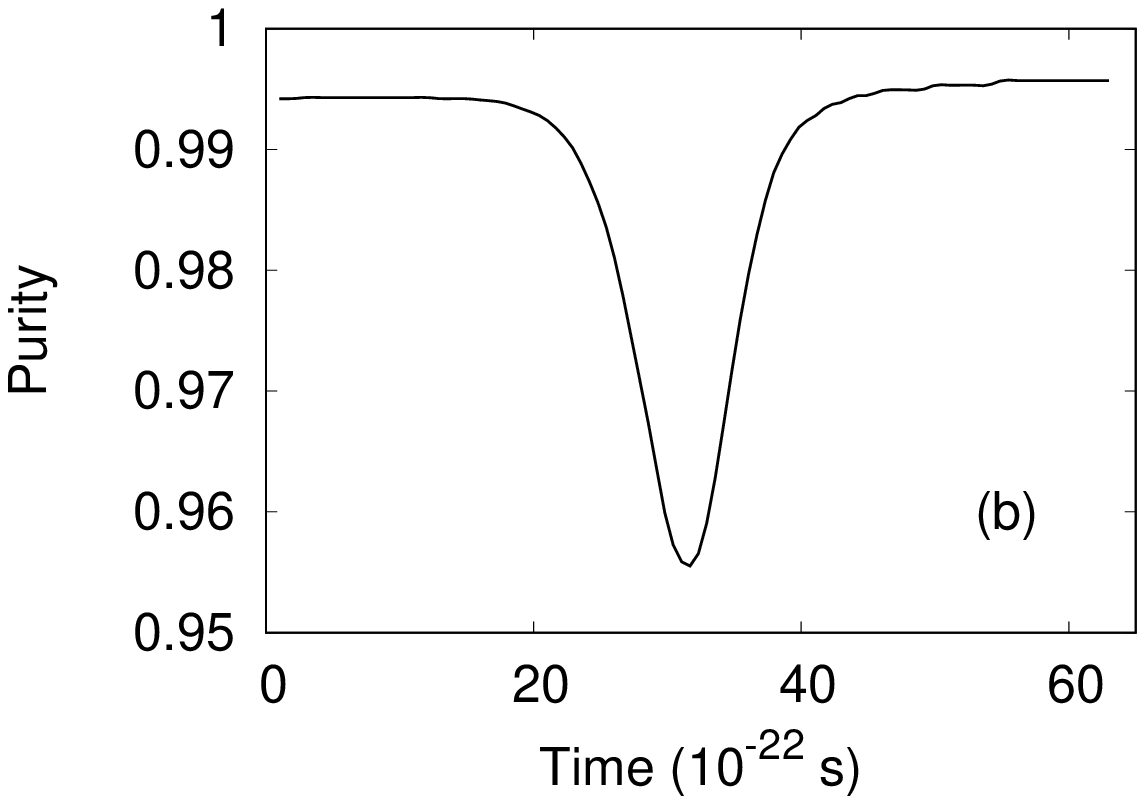}
\label{fig:entropy}} 
 \caption{The dynamics of energy dissipation and quantum coherence in the collision scenario shown in Fig. \ref{fig:globfig}. (a) Energy loss (thick solid red line) and entropy production (thin solid black line) increase as the nuclei approach their Coulomb barrier, and reach a plateau when the nuclei re-separate. The dotted black line shows the average internuclear radius of the reduced system. (b) The purity of the reduced-sytem density matrix decreases (decoherence) when the nuclei are near the Coulomb barrier. There is coherence resurgence (purity increases and entropy decreases) when the nuclei re-separate as the interaction with the fusion environment diminishes. Asymptotically, the elastic and inelastic channels evolve in a coherent superposition.}
 \label{fig:coherence_dynamics}
\end{figure}

One useful feature of this method is the ability to follow the dynamics in time and understand what is happening at each time step of the propagation. Fig. \ref{fig:globfig} shows the population of the radial grid basis states over time for an initial Coulomb wave packet with an average energy $E_0=60$ MeV. The change of population of both the ground state and the $3^{-}$ excited state of $^{144}$Sm due to the radial coupling between these states can be observed.

Fig. \ref{fig:coherence_dynamics} displays the dynamics of (i) energy dissipation (thick solid red line), entropy production (thin solid black line) [panel (a)], and (ii) quantum coherence (solid line) [panel (b)] for the collision scenario shown in Fig. \ref{fig:globfig}. The energy loss is determined by the change of the average energy of the reduced system relative to its initial value, i.e., $E_0 - \textnormal{Tr}[\hat{H}_S \rho_S(t)]$. It is interesting to observe that the transient, strong interaction of the reduced system with the fusion environment makes the dynamics dissipative and decoherent (the entropy increases and the purity of the reduced-system density matrix decreases). When the nuclei re-separate, there is a revival of quantum coherence and, asymptotically, the elastic and inelastic channels move in a coherent superposition. Does the transient decoherent phase of the collision affect the fusion probability?   

\begin{figure}[htb!]
 \centering
 \includegraphics[scale=0.85]{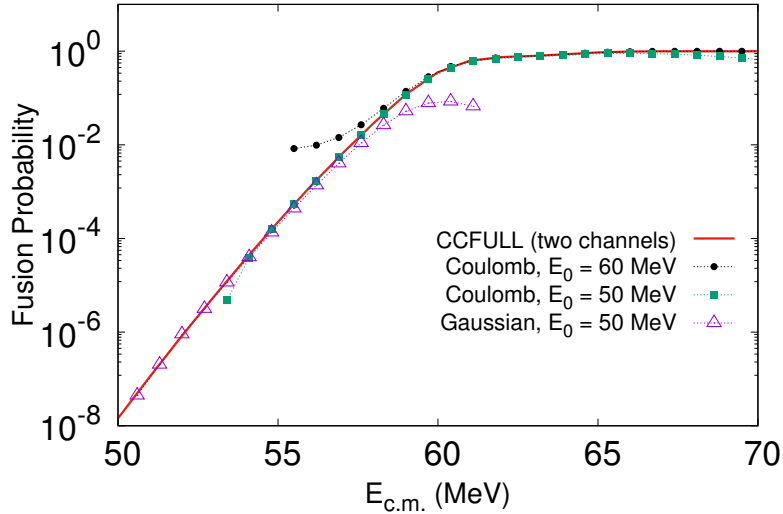}
 \caption{\label{fig:fusion_prob} The energy resolved fusion probabilities from the CCDM method for different initial wave packets for a head-on collision of $^{16}$O + $^{144}$Sm with various mean energies. These probabilities are compared with those from CCFULL \cite{ccfull}. The CCFULL results (red solid line) are reproduced by the CCDM method (symbols).}
\end{figure}

Figure \ref{fig:fusion_prob} shows the fusion probabilities calculated using optimal conditions for the initial Gaussian and Coulomb wave packets for the test case collision $^{16}$O + $^{144}$Sm. The height of the Coulomb barrier between these two nuclei is $V_B=61.1$ MeV. 
The Gaussian wave packet calculations clearly provide the best results in terms of energy-resolved fusion probability at deep sub-barrier energies ($E_{c.m.}/V_B < 0.9$), reaching an order of magnitude of $10^{-8}$ (purple triangles). Having accurate values of fusion probability for low-energy fusion reactions allow us to extend the scope of our calculations and will allow us to validate interesting effects at these low fusion probability regions. This could be the case when, for example, using a plasma environment.

The Coulomb wave packets can still achieve values of $\sim 10^{-6}$ (green solid squares), however the broadness of these wave packets in momentum space allows a better global description of energy values below, near and above the barrier ($E_{c.m.}/V_B > 0.9$), as also shown in Ref. \cite{terry2021}. This could be useful as less computational time is needed if the lowest penetration values are not so important. The physically meaningful values of the CCDM fusion probability are the invariant values when the parameters of the initial wave packet are changed \cite{boselli2015,alexis2018,terry2019}. The probabilities that are out of this trend are determined with a very small, initial weighting factor in the window function and are numerically inaccurate. 

The very good agreement between the CCDM and CCFULL fusion calculations demonstrates the reliability of the CCDM approach. Since CCFULL does not include decoherence, these results indicate that the transient quantum decoherence caused by the fusion environment localised in the nucleus-nucleus potential pocket does not affect the fusion probability. Quantum decoherence is a dynamical phenomenon \cite{schlosshauer2007} which cannot be described by a static coupled-channels model such as CCFULL. 
Dynamical and static coupled-channels calculations based on the Schr\"odinger equation provide the same (asymptotic) observables, as shown in Ref. \cite{terry2021}. The same is expected to happen using a semi-classical dynamical coupled-channels model \cite{Laura2021} at above-barrier incident energies. 
In such semi-classical calculations, without complex potentials, the occupation probability of direct reaction channels oscillates in time at internuclear radii around the distance of minimal approach (e.g., see Fig. 2 in Ref. \cite{Laura2021}). These oscillations are due to the reversible couplings among the reaction channels which have not yet achieved equilibrium in their occupation probabilities. The reversible couplings do not affect the coherent superposition of intrinsic energy states. In the present work, the change of purity in Fig. 3(b) is caused by the effect of the irreversible coupling between the fusion environment and the reduced system. Without the fusion environment, the dynamics of the CCDM model would be Hamiltonian, preserving energy, entropy and the purity of the reduced density matrix.    

\section{Conclusions}
\label{section:conclusion}

We have successfully shown that the CCDM approach can be used for describing low-energy nuclear fusion reactions. The window operator has been applied to a density matrix for energy resolving purposes, which in this case, is crucial for validating the treatment of fusion using the CCDM method. It is shown that the CCDM and TISE calculations of fusion are in very good agreement with each other. Unexpectedly, the transient decoherence within the fusion pocket does not affect the fusion probability \cite{Dasgupta2007}. The CCDM approach has the advantage of allowing us to study quantities throughout the time propagation of the density matrix, such as coherence dynamics, energy dissipation and entropy production. Most importantly, we have built a foundation for the study of environmental effects on a fusion reaction, via the open quantum systems methodology. The next step is to introduce a physical, plasma environment which is present in all stellar fusion reactions. This will be implemented in a similar manner as the fusion environment, using specifically chosen Lindblad operators. Achieving accurate fusion probabilities well-below the Coulomb barrier will allow us to extend our study since plasma interactions are more likely to affect sub-barrier fusion. Further questions, such as the experimentally observed hindrance of fusion at sub-barrier energies \cite{hindrance2021} and dissipative multi-nucleon transfer reactions at sub-barrier energies \cite{mnt2020}, may also be addressed.

\textbf{Acknowledgement.} This work has received funding from the Leverhulme Trust (UK) under Grant No. RPG-2019-325.





\begin{thebibliography}{00}


\bibitem[Yakovlev(1989)]{Yakovlev1989} D.G. Yakovlev, D.A. Shalybkov, \emph{Degenerate cores of white dwarfs and envelopes of neutron stars: Thermodynamics and plasma screening in thermonuclear reactions}, Astrophys. Space Phys. 7 (1989) 311.


\bibitem[Harston(1999)]{Harston1999} M. Harston, J.F. Chemin, \emph{Mechanisms of nuclear excitation in plasmas}, Phys. Rev. C 59 (1999) 2462.


\bibitem[Adriana(2006)]{adriana2006} A. P\'alffy, W. Scheid, Z. Harman, \emph{Theory of nuclear excitation by electron capture for heavy ions}, Phys. Rev. A 73 (2006) 012715.

\bibitem[Gosselin(2009)]{Gosselin2009} G. Gosselin, N. Pillet, V. M\'eot, P. Morel, A.Ya. Dzyublik, \emph{Nuclear transition induced by low-energy unscreened electron inelastic scattering}, Phys. Rev. C 79 (2009) 014604.


\bibitem[breuer(2002)]{breuer2002} H. Breuer, F. Petruccione, \emph{The Theory of Open Quantum Systems}, Oxford University Press, 2002.

\bibitem[rotter(2015)]{rotter2015} I. Rotter, J.P. Bird, \emph{A review of progress in the physics of open quantum systems: theory and experiment}, Rep. Prog. Phys. 78 (2015) 114001.

\bibitem[sandulescua(1987)]{sandulescua1987} A. Sandulescu, H.Scutaru, \emph{Open quantum systems and the damping of collective modes in deep inelastic collisions}, Ann. Phys. (NY) 173 (1987) 277.






\bibitem[vazgen(2016)]{vazgen2016} V.V. Sargsyan, Z. Kanokov, G.G. Adamian, N.V. Antonenko, \emph{Application of the theory of open quantum systems to nuclear physics problems}, Phys. Part. Nucl. 47 (2016) 157.

\bibitem[Dasgupta(2007)]{Dasgupta2007} M. Dasgupta, D.J. Hinde, A. Diaz-Torres, B. Bouriquet, C.I. Low, G. Milburn, J.O. Newton, \emph{Beyond the coherent coupled channels description of nuclear fusion}, Phys. Rev. Lett. 99 (2007) 192701.



\bibitem[Lindblad1976(1976)]{Lindblad1976} G. Lindblad, \emph{On the generators of quantum dynamical semigroups}, Comm. Math. Phys. 48 (1976) 119.


\bibitem[ pesce(1997)]{pesce1997} L. Pesce, P. Saalfrank, \emph{``Free" nuclear density propagation in two dimensions: the coupled-channel density matrix method and its application to inelastic molecule-surface scattering}, Chem. Phys. 219 (1997) 43.

\bibitem[ alexis2008(2008)]{alexis2008} A. Diaz-Torres, D.J. Hinde, M. Dasgupta, G.J. Milburn, J.A. Tostevin, \emph{Dissipative quantum dynamics in low-energy collisions of complex nuclei}, Phys. Rev. C 78 (2008)064604.  

\bibitem[ alexis2010(2010)]{alexis2010} A. Diaz-Torres, \emph{Coupled-channels density-matrix approach to low-energy nuclear collision dynamics: A technique for quantifying quantum decoherence effects on reaction observables}, Phys. Rev. C 82 (2010) 054617.




\bibitem[ ccfull(1999)]{ccfull} K. Hagino, N. Rowley, A. Kruppa, \emph{A programme for coupled-channel calculations with all order couplings for heavy-ion fusion reactions}, Comp. Phys. Comm. 123 (1999) 143.

\bibitem[ schafer(1991)]{schafer1991} K. Schafer, \emph{The energy analysis of time-dependent numerical wave functions}, Comp. Phys. Comm. 63 (1991) 427.

\bibitem[ boselli(2015)]{boselli2015} M. Boselli, A. Diaz-Torres, \emph{Quantifying low-energy fusion dynamics of weakly bound nuclei from a time-dependent quantum perspective}, Phys. Rev. C 92 (2015) 044610.

\bibitem[ alexis(2018)]{alexis2018} A. Diaz-Torres, M. Wiescher, \emph{Characterizing the astrophysical S factor for $^{12}$C + $^{12}$C fusion with wave-packet dynamics}, Phys. Rev. C 97 (2018) 055802.

\bibitem[ terry(2019)]{terry2019} T. Vockerodt, A. Diaz-Torres, \emph{Describing heavy-ion fusion with quantum coupled-channels wave-packet dynamics}, Phys. Rev. C 100 (2019) 034606.


\bibitem[kosloff(2019)]{kosloff2019} R. Kosloff, \emph{Quantum thermodynamics and open-systems modeling}, J. Chem. Phys. 150 (2019) 204105.

\bibitem[Balantekin(1998)]{Balantekin1998} A.B. Balantekin, N. Takigawa, \emph{Quantum tunneling in nuclear fusion}, Rev. Mod. Phys. 70 (1998) 77.

\bibitem[Hagino(2012)]{Hagino2012} K. Hagino, N. Takigawa, \emph{Subbarrier fusion reactions and many-particle quantum tunneling}, Prog. Theor. Phys. 128 (2012) 1061. 


\bibitem[thompson(2001)]{thompson2001} I.J. Thompson, \emph{Methods of direct reaction theories}, in \emph{Scattering}, E.R. Pike and P.C. Sabatier (eds), Academic Press, 2001, pp. 1360-1372.

\bibitem[bertlmann(2006)]{bertlmann2006} R.A. Bertlmann, W. Grimus, B.C. Hiesmayr, \emph{Open-quantum-system formulation of particle decay}, Phys. Rev. A 73 (2006) 054101.

\bibitem[thompson(2004)]{thompson2004} I.J. Thompson, A. Diaz-Torres, \emph{Modelling effects of halo breakup on fusion}, Prog. Theor. Phys. Suppl. 154 (2004) 69.


\bibitem[thompson(1986)]{thompson1986} A.R. Barnett, \emph{COULFG: Coulomb and Bessel functions and their derivatives, for real arguments, by Steed's method}, Comp. Phys. Comm. 27 (1982) 147.

\bibitem[ terry(2021)]{terry2021} T. Vockerodt, A. Diaz-Torres, \emph{Calculating the S-matrix of low-energy heavy-ion collisions using quantum coupled-channels wave-packet dynamics}, Phys. Rev. C 104 (2021) 064601.

\bibitem[huisinga(1999)]{huisinga1999} W. Huisinga, L. Pesce, R. Kosloff, P. Saalfrank, \emph{Faber and Newton polynomial integrators for open-system density matrix propagation}, J. Chem. Phys. 110 (1999) 5538.

\bibitem[Saalfrank(1995)]{Saalfrank1995} C. Scheurer, P. Saalfrank, \emph{Density matrix model for hydrogen transfer in the benzoic acid dimer}, Chem. Phys. Lett. 245 (1995) 201.

\bibitem[lenton(2021)]{lenton2021} J. Lenton, I. Lee, A. Diaz-Torres, \emph{Quantum dynamics of a nucleon in the Fermi accelerator}, Ann. Phys. (NY) 434 (2021) 168624.

\bibitem[schlosshauer(2007)]{schlosshauer2007} M. Schlosshauer, \emph{Decoherence and The Quantum-to-Classical Transition}, Springer Verlag, 2007. 


\bibitem[Laura(2021)]{Laura2021} L. Moschini, A. Diaz-Torres, \emph{Tracing the dynamical interplay of low-energy reaction processes of exotic nuclei using a two-center molecular continuum}, Phys. Lett. B 820 (2021) 136513. 


\bibitem[hindrance(2021)]{hindrance2021} C.L. Jiang, B.B. Back, K.E. Rehm, K. Hagino, G. Montagnoli, A.M. Stefanini, \emph{Heavy-ion fusion reactions at extreme sub-barrier energies}, Eur. Phys. J. A 57 (2021) 235.  

\bibitem[mnt(2020)]{mnt2020} G.G. Adamian, N.V. Antonenko, A. Diaz-Torres, S. Heinz, \emph{How to extend the chart of nuclides?}, Eur. Phys. J. A 56 (2020) 47. 


\end{thebibliography}



\end{document}